\documentclass[11pt]{article}

\usepackage{dcolumn, url}
\usepackage{bm}
\usepackage{amssymb, amsmath, amscd, xypic}
\usepackage{graphicx}
\usepackage[latin1]{inputenc}
\usepackage{mathrsfs}
\usepackage{epsfig, epic, eepic}
\usepackage{tikz}
\usepackage{epstopdf}
\usepackage{bbm,dsfont}
\usepackage{cite}
\usepackage{setspace}

\newcommand{\state}[1]{\left\vert #1 \right\rangle}

\usepackage{color}

\newcommand{\id}{\mathbbm{1}}

\newtheorem{theorem}{Theorem}[section]

\newtheorem{definition}[theorem]{Definition}

\begin{document}


\newcommand{\calA}{{\cal A}}
\newcommand{\calB}{{\cal B}}
\newcommand{\calF}{{\cal F}}
\newcommand{\calG}{{\cal G}}
\newcommand{\calR}{{\cal R}}
\newcommand{\calZ}{{\cal Z}}

\newcommand{\M}{{\mathcal M}}
\newcommand{\X}{{\mathcal X}}

\newcommand{\A}{{\mathscr{A}}} 
\newcommand{\G}{{\mathscr{G}}}
\newcommand{\Osheaf}{{\mathscr{O}}}

\newcommand{\N}{{\mathbb N}}
\newcommand{\Z}{{\mathbb Z}}
\newcommand{\Q}{{\mathbb Q}}
\newcommand{\R}{{\mathbb R}}
\newcommand{\CC}{{\mathbb C}}

\newcommand{\F}{{\mathbb F}}

\newcommand{\K}{{\mathbb K}}
\newcommand{\kk}{{\mathrm k}}

\def\J{\mathrm{J}}
\def\catC{{\bf \mathrm{C}}}
\def\x{\mathrm{x}}
\def\a{\mathrm{a}}
\def\d{\mathrm{d}}
\def\Bi{\mathrm{Bi}}
\def\op{\mathrm{op}}
\def\res{\mathrm{res}}
\def\span{\mathrm{span}}

\newcommand{\C}{{\bf C}}
\newcommand{\Objects}{{\bf Objects}}
\newcommand{\Arrows}{{\bf Arrows}}
\newcommand{\Sets}{{\bf Sets}}

\newenvironment{menumerate}{%
    \renewcommand{\theenumi}{\roman{enumi}}%
    \renewcommand{\labelenumi}{\rm(\theenumi)}%
    \begin{enumerate}} {\end{enumerate}}
     
\newenvironment{system}[1][]%
	{\begin{eqnarray} #1 \left\{ \begin{array}{lll}}%
	{\end{array} \right. \end{eqnarray}}

\newenvironment{meqnarray}%
	{\begin{eqnarray}  \begin{array}{rcl}}%
	{\end{array}  \end{eqnarray}}

\newenvironment{marray}%
	{\\ \begin{tabular}{ll}}
	{\end{tabular}\\}

\newenvironment{program}[1]%
	{\begin{center} \hrulefill \quad {\sf #1} \quad \hrulefill \\[8pt]
		\begin{minipage}{0.90\linewidth}}
	{\end{minipage} \end{center} \hrule \vspace{2pt} \hrule}

\newcommand{\entrylabel}[1]{\mbox{\textsf{#1:}}\hfil}
\newenvironment{entry}
   {\begin{list}{}%
   	{\renewcommand{\makelabel}{\entrylabel}%
   	  \setlength{\labelwidth}{40pt}%
   	  \setlength{\leftmargin}{\labelwidth + \labelsep}%
   	}%
   }%
   {\end{list}}

\newenvironment{Pmatrix}
        {$ \left( \!\! \begin{array}{rr} } 
        {\end{array} \!\! \right) $}

\newcommand{\into}{{\;\rightarrow\;}}
\newcommand{\Hom}{\operatorname{Hom}}
\newcommand{\pref}[1]{(\ref{#1})}
\newcommand{\frakM}{\mathfrak{M}}
\newcommand{\card}{\operatorname{card}}
\newcommand{\Spec }{\mathrm{ Spec\,}}
\newcommand{\p}{\mathfrak{p}}

{
}

\title{The many mathematical faces of Mermin's proof of the Kochen-Specker theorem}

\author{
  Leon Loveridge$^{a,}$\thanks{leon.loveridge@cs.ox.ac.uk} {\ and} 
Raouf Dridi$^{b,}$\thanks{dridi.raouf@gmail.com}\\  
{\small\sl a. Department of Computer Science, University of Oxford,}\\ 
{\small\sl Wolfson Building, Parks Rd, Oxford, UK OX1 3QD}\\   
{\small\sl b. Department of Mathematics, University of British Columbia,}\\ 
{\small\sl Vancouver,  BC V6T 1Z2, Canada}
}

\date{\today}
 \maketitle
\begin{abstract}
Mermin's simple ``pentagram" proof of the Kochen-Specker theorem is examined from various perspectives. We emphasise the many mathematical structures intimately related to Kochen-Specker proofs, ranging through functional analysis, sheaf theory and topos theory, Coxeter groups and algebraic geometry.
Some novel results are presented along the way.
\end{abstract}









\section{Introduction}

The assortment of results collectively referred to as the \emph{Kochen-Specker theorem}, discovered by Bell \cite{bell} and Kochen and Specker \cite{KS1} in the 1960s, still occupies a prominent position in the foundations of quantum mechanics. The physical and philosophical content of these theorems expound essential differences between classical and quantum mechanics, 
{\it grosso modo} informing us that the mathematical structure of quantum theory is incompatible with a realist interpretation of the theory in the spirit of classical statistical mechanics. 

There now exists a vast literature on many aspects of the Kochen-Specker theorem. On the mathematical side this  includes ``small"/efficient proofs (e.g., \cite{peres2, penrose1} for proofs in $\mathbb{R}^3$, the record there for the smallest number of rays being held by Conway and Kochen---see \cite{peres1}), general graph theoretic treatments \cite{csw1, csw2, afls1}, generalisations to unsharp observables (e.g., \cite{spek1, paul1, cab1}), preparation and transformation contextuality \cite{spek1}, operator algebras \cite{dor1}, (co-)sheaf and\linebreak (co-)presheaf \cite{ib1,ib2,ib3,ib4,hls1,abram1} approaches and geometry and cohomology \cite{abram2, abrampara}, ``logical" Bell inequalities \cite{ah1}, topos theory \cite{id1, ldr1, hls1},  and non-contextual inequalities \cite{kly1,csw1} akin in style to the Bell inequality \cite{bell1, chsh1}, whose violation detects contextuality, to name but a few. 
Contextuality has also recently been pinpointed as a resource for quantum
computation \cite{rau1,rau2,hwve1}.

In this paper we use Mermin's simple ``pentagram" proof \cite{Mermex} of the Kochen-Specker theorem in Hilbert space dimension $8$, along with a corresponding state-dependent version, to emphasise the many
mathematical structures intimately related to Kochen-Specker proofs.
This functions as an introduction to the subject of Kochen-Specker proofs as well as a 
survey of the many mathematical viewpoints that now exist on the Kochen-Specker theorem.

After providing some background and motivating the problem that Bell and Kochen and Specker addressed, we give Mermin's state-independent proof based on the non-existence of a valuation for 10-qubit observables, as well as the corresponding state-dependent version. This gives rise to colouring proofs, based on the impossibility of assigning values to projections/rays arising from the spectral resolution of the given observables. We mention another general form of Kochen-Specker scenario called ``all-versus-nothing" arguments \cite{abrampara} of which the Mermin system is an example. We also present a result due to Clifton \cite{clif1}
showing that Mermin's proof may be adapted to the case of position-momentum contextuality for three degrees of freedom. An algebraic perspective due to D\"{o}ring \cite{dor1} is then presented,
before a discussion of Coxeter groups, $E_8$ in particular, which is seen to naturally arise from the
Mermin system.

Three sheaf-theoretic formulations are then given, based on the work of Isham and collaborators and focussing on the so-called spectral presheaf (e.g., \cite{id1}), followed by a covariant approach of Heunen, Landsman, and Spitters \cite{hls1}, and finally an operational approach initiated by Abramsky and Brandenburger \cite{abram1}. A novel algebraic geometric framework is then provided, followed by a short section providing some conclusions and avenues for further investigation.

\section{Background on the Kochen-Specker Theorem}
A basic question in the foundations of quantum mechanics is whether the probabilistic structure
of the orthodox formalism is a reflection of a fundamental property of the quantum world, or comes about as an effective description arising from the theorist's lack of detail about the inner workings of quantum phenomena. 

If it is the latter, it is compelling to believe that quantum mechanical observables should have values prior to, and independently of, measurement. Although these values cannot, in general, be realised by a quantum state, they should, in principle, be realised by a ``microstate". In Newtonian physics,
an observable quantity is represented by a\linebreak (Borel/continuous/smooth) function $f:T^*\mathcal{M} \to \mathbb{R}$ on the phase space $T^*\mathcal{M}$. The value $\nu_s(f)$ of $f$ in a state $s \in T^*\mathcal{M}$ is given by $\nu_s(f)=f(s)$. Therefore, the possible values of an observable $f$ are the spectral values of $f$ in the sense of $C^*$-algebras. Moreover, given a function 
$h:\mathbb{R} \to \mathbb{R}$, the observable $h(f) :=h \circ f : T^*\mathcal{M} \to \mathbb{R}$
has value $\nu_s(h \circ f) = h(f(s))$ in the state $s \in T^*\mathcal{M}$.

It is then natural to look for a ``hidden state" underlying the quantum description that behaves analogously to a state
in classical physics, and specifically that values quantum observables in an appropriate way.
With $\mathcal{H}$ a complex separable Hilbert space and $\mathcal{L(H)}$ the collection of bounded linear operators (to be viewed as either a von Neumann or $C^*$-algebra), the hidden state 
would then be given as a valuation obeying analogous conditions to the classical case, namely
$\nu : \mathcal{L(H)} \to \mathbb{C}$ such that for any self-adjoint element $a \in \mathcal{L(H)}$ and function
$f: \mathbb{R} \to \mathbb{R}$,

\begin{enumerate}
\item $\nu(a) \in spec(a)$ (SPEC),
\item $\nu (f(a)) = f(\nu(a))$ (FUNC),
\end{enumerate}
where condition (2) (FUNC) is understood in the sense of the appropriate functional calculus. An immediate consequence of FUNC is that $\nu$ is linear and multiplicative on commuting operators.
Alternatively, one can say that $\nu$ is a multiplicative linear
functional on $C^*(a)$ (with the natural extension to complex combinations), and is therefore
a character of $C^*(a)$. In a finite dimensional Hilbert space, the collection of all such characters are in one-to-one correspondence with the set of non-zero eigenvalues of $a$ and, therefore, characters are generic as valuations.

We also note that $\nu$ can be extended naturally to the whole of $\mathcal{L(H)}$ and FUNC
to complex-valued functions of the appropriate type.

\section{Mermin's Pentagram: Combinatoric Proofs}

\subsection{Colouring and parity proofs}
Mermin provided a simple construction for ruling out a valuation $\nu$ on a collection of ten
self-adjoint operators in $\mathbb{C}^8$, arranged in five \emph{contexts}, given here as collections of commuting observables. With
$\sigma_{x,z} \in M_2(\mathbb{C})$ denoting the Pauli operators and writing $X_1 := \sigma_x \otimes \id \otimes \id$, etc., applying FUNC (and for the final equality, also SPEC) yields

\begin{eqnarray}\label{eq:mermcon}
\nu(X_1X_2X_3) = \nu(X_1)\nu(X_2)\nu(X_3),\label{eq:1}\\
\nu(X_1Z_2Z_3) = \nu(X_1)\nu(Z_2)\nu(Z_3),\label{eq:2}\\
\nu(Z_1X_2Z_3) = \nu(Z_1) \nu (X_2) \nu (Z_3),\label{eq:3}\\
\nu(Z_1Z_2X_3) = \nu(Z_1)\nu (Z_2) \nu (X_3),\label{eq:4} \\
\nu(X_1X_2X_3)\nu(X_1Z_2Z_3) \nu(Z_1X_2Z_3) \nu(Z_1Z_2X_3)=-1 \label{eq:5};
\end{eqnarray}
for the final equation each observable 
appearing as an argument in $\nu$ commutes with all others inside a 
$\nu(\cdot)$, and we exploit the anticommutativity of $X_i$ and $Z_i$.
On the other hand, taking products of the left-hand sides and right-hand sides of \eqref{eq:1}--\eqref{eq:4} gives
\begin{equation}\label{eq:5-}
\nu(X_1X_2X_3)\nu(X_1Z_2Z_3) \nu(Z_1X_2Z_3) \nu(Z_1Z_2X_3)=1
\end{equation} by SPEC, and therefore there exists no context-independent valuation $\nu$ compatible
with the above constraints, yielding a combinatoric obstruction to a hidden state acting as a valuation. This contradiction
constitutes a so-called \emph{parity proof} of the Kochen-Specker theorem.

\emph{Colouring proofs} can also be extracted from the Mermin system in more than one way;
Kerhaghan and Peres \cite{kpe1}, Waegell and Aravind \cite{arw1}, and Toh \cite{toh1} have each given an example.
Kernaghan and Peres note that each context generates 8 mutually orthogonal vectors, given as the simultaneous eigenvectors of the (commuting) observables appearing in the context. Following Waegell and Aravind \cite{arw1}, we may form the rank-one
projections $P[v_i]\equiv P_i$, and given that for the first context, for example,
 $\sum_{i=1}^8 P_i = \id$, we have $\sum_{i=1}^8 \nu(P_i)=1$. Therefore, precisely
 one $P_i$ must have value $\nu(P_i)=1$, with all others having a value of zero. This amounts
 to colouring precisely one vector/ray out of a complete orthogonal set green, for instance, and all of the others red. The assumption of non-contextuality manifests as the requirement that the given colour must be independent of the context with which the projection is associated (since the contexts have non-zero intersections). It is found that such a colouring is impossible, yielding another combinatoric proof. 
 
To conclude this subsection, we also mention state-dependent versions of the Kochen-Specker theorem
based on the Mermin system. Equation \eqref{eq:5} is consistent with 
$\nu(X_1X_2X_3) = 1$, $\nu(X_1Z_2Z_3) = -1$, $\nu(Z_1X_2Z_3) = -1$, and $\nu(Z_1Z_2X_3) = -1$, yielding
the equations
\begin{eqnarray}\
\nu(X_1)\nu(X_2)\nu(X_3)=1, \label{eq:MS1}\\
\nu(X_1)\nu(Z_2)\nu(Z_3)=-1, \label{eq:MS2}\\
\nu(Z_1) \nu(X_2)\nu(Z_3) = -1, \label{eq:MS3} \\
\nu (Z_1) \nu(Z_2) \nu(X_3) = -1. \label{eq:MS4}
\end{eqnarray}

Again, by taking products of the left-hand side and the right-hand side, a contradiction arises.
This is clearly weaker than the contradiction implied by equations \eqref{eq:1}--\eqref{eq:5}.
The state-dependence comes from observing that the numbers on the right-hand side are eigenvalues of the products $X_1X_2X_3$, $X_1Z_2Z_3$, etc., in the Greenberger-Horne-Zeilinger (GHZ) state 
$\Psi = \frac{1}{\sqrt{2}}\left(\state{000}+\state{111} \right)$ \cite{GHZ} . Thus the contradiction may be understood as the impossibility of the given observables having values compatible with the quantum predictions in
a specific state, in this case the GHZ state $\state{\Psi}$.

The Hasse diagram illustrating the state-dependent Mermin poset $\mathcal{P}$ is as follows:

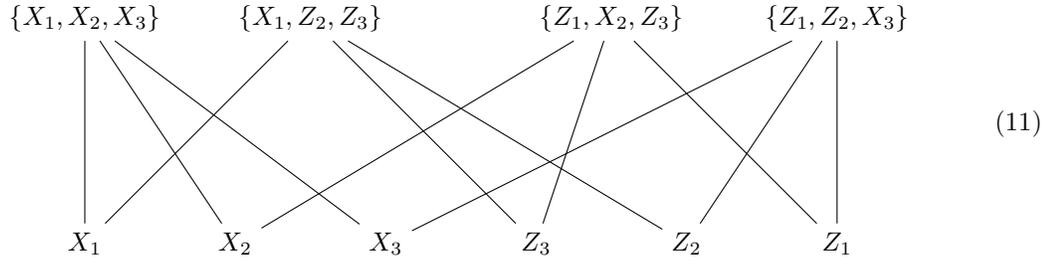
\begin{small}
\begin{center}
\begin{equation}\label{d:ghz} 
\begin{tikzpicture}[baseline=(current  bounding  box.center)]

  \node (g1) at (-5,3) {$\{X_1,X_2,X_3\}$};
  \node (g2) at (-2,3) {$\{X_1, Z_2, Z_3 \}$};
  \node (g3) at (2,3) {$\{Z_1, X_2, Z_3\}$};
   \node (g4) at (5,3) {$\{Z_1, Z_2, X_3 \}$};
 
  \node (x1)at (-5,0) {${X_1}$};
  \node (x2) at (-3,0) {${X_2}$};
  \node (x3) at (-1,0) {${X_3}$};
 \node (y3) at (1,0) {$Z_3$};
  \node (y2)at (3,0) {$Z_2$};
  \node (y1)at (5,0) {$Z_1$}; 

 \draw  (g1)--(x1);
  \draw  (g1)--(x2) ;
  \draw  (g1)--(x3) ;

 \draw  (g2)--(x1) ;
  \draw  (g2)--(y2);
  \draw  (g2)--(y3);

 \draw  (g3)--(x2);
  \draw  (g3)--(y3) ;
  \draw  (g3)--(y1);

  \draw  (g4)--(y2) ;
  \draw  (g4)--(x3);
  \draw  (g4)--(y1);

\end{tikzpicture}
\end{equation}
\end{center}
\end{small}

\subsection{``All-versus-nothing" arguments}

Mermin \cite{Mermex} dubbed the contradiction obtained from the GHZ state (i.e., \eqref{eq:MS1}--\eqref{eq:MS4}) 
an ``all-versus-nothing" (AvN) argument. Similar strong forms of contextuality (to be discussed further in Section \ref{subsec:Abram}) have appeared in the literature, and have been generalised in \cite{abrampara}, where a general form of AvN argument is provided based on $mod-n$ linear
equations and is independent of the quantum formalism.

A rich source of examples of quantum AvN arguments is provided in the setting of stabiliser quantum mechanics. $\mathcal{P}_n$ denotes the $n$-qubit Pauli group,
i.e., products of $\mathcal{P}_1 = \left\langle j X,  jY , j Z, j \id\right\rangle$, with 
$j \in \{\pm 1, \pm i \}$. Then, an AvN argument, in direct analogy with the GHZ case already considered, is a system of local equations of parity type which have no global solution.

The main theorem on this subject in \cite{abrampara} shows that any 
``AvN triple" gives rise to an AvN argument.

\begin{definition} 
An AvN triple $\left\langle e,f,g \right\rangle$ in $\mathcal{P}_n$ is a triple $\left\langle e,f,g \right\rangle$ of elements of $\mathcal{P}_n$ with global phases +1 which pairwise commute, and which satisfy the following conditions:
\begin{enumerate}
\item For each $i \in \{1,...,n\}$ at least two of $e_i$, $f_i$, $g_i$ are equal.
\item The number of $i$ such that $e_i = g_i = f_i$, all distinct from $\id$, is odd.
\end{enumerate}

\end{definition}

Then,

\begin{theorem}
Let S be the subgroup of $\mathcal{P}_n$ generated by an AvN triple, and $V_S$ the subspace stabilised by $S$. For every state $\Psi$ in $V_S$, the empirical model (a precise definition is given in \ref{subsec:Abram}) realised by $\Psi$ under the Pauli measurements admits an all-versus-nothing argument.
\end{theorem} 
Here, ``empirical model" refers to the probability table arising from joint distributions defined by $\Psi$ and the observables appearing in each context.

\subsection{Position-Momentum contextuality for three spacetime degrees of freedom}

Clifton \cite{clif1} has observed that the Mermin proof encapsulated in equations
\eqref{eq:1}--\eqref{eq:5} may be adapted to provide a proof of contextuality based on the position and momentum of a spinless particle in three spacetime dimensions. He considers the Weyl form of the canonical commutation relation (CCR) in $\mathcal{H} \equiv L^2(\mathbb{R}^3)$. With $Q$ and $P$  the usual position and momentum operators acting in (some dense domain of)
$\mathcal{H}$ (i.e., $Q=(Q_1, Q_2, Q_3)$, etc.), $p,q \in \mathbb{R}^3$, and writing $U(p)\equiv e^{ip.Q}$ and
$V(q) \equiv e^{iq.P}$, the CCR takes the form
\begin{equation}\label{eq:Weyl}
U(p)V(q)=e^{-ip.q}V(q)U(p).
\end{equation}
The component operators $U(p_i) \equiv e^{ip_iQ_i}$ and $V_i = e^{i q_i P_i}$ satisfy the same
form of CCR as \eqref{eq:Weyl} for each $i$; furthermore,
\begin{equation}\label{eq:com}
[U(p_i),V(q_i)]=0 \text{~whenever~} p_iq_i = 2ni\pi,
\end{equation}
and, with $[\cdot,\cdot]_+$ denoting the anticommutator,
\begin{equation}\label{eq:acom}
[U(p_i),V(q_i)]_+=0 \text{~whenever~} p_iq_i = (2n+1)i\pi.
\end{equation}
Then, as Clifton observed, one may consider once more a valuation $\nu$, defined on self-adjoint
operators in $\mathcal{L(H)}$ and naturally extended to all bounded operators in the obvious way.

The commutativity and anticommutativity displayed by equations \eqref{eq:com} and \eqref{eq:acom}
suggest that one may find suitable values of $p_i,q_i$ to reproduce the Mermin proof, and this is precisely what is done in \cite{clif1}, where Clifton chooses $p_iq_i=(2n+1)\pi$ for all $i$.
The Mermin contradiction stemming from equations \eqref{eq:1}--\eqref{eq:5} relies on the fact that the operators in question are involutions, and therefore that $\nu(X_i^2) = \nu(Z_i^2)=1$. Such a property is, however, lost for $U(p_i)$ and $V(q_i)$. Nevertheless, with $U(-q_i)$, the inverse of
$U(q_i)$, etc., the equations 
\begin{eqnarray}
\nu \left( U(-p_1)U(-p_2)U(-p_3)) = \nu(U(-p_1))\nu(U(-p_2))\nu(U(-p_3) \right),\label{eq:1''}\\
\nu(V(q_1)V(q_2)U(p_3)) = \nu(V(q_1))\nu(V(q_2))\nu(U(p_3)),\label{eq:2''}\\
\nu(V(-q_1)U(p_2)V(-q_3)) = \nu(V(-q_1))\nu(U(p_2))\nu(V(-q_3)),\label{eq:3''}\\
\nu(U(p_1)V(-q_2)V(-q_3)) = \nu(U(p_1))\nu(V(-q_2))\nu(V(-q_3)),\label{eq:4''}
\end{eqnarray}
hold by FUNC.
Precisely akin to the Mermin proof, taking products of the right-hand sides and left-hand sides therefore gives 
\begin{multline}
\nu(U(-p_1)U(-p_2)U(-p_3))\nu(V(q_1)V(q_2)U(p_3))\nu(V(-q_1)U(-p_2)V(-q_3)) \\ \nu(U(p_1)V(-q_2)V(-q_3))=1,
\end{multline}
and, in direct analogy to Mermin, noticing that each operator appearing as an argument of $\nu$ commutes with all others appearing in $\nu$, the anticommutativity yields 
\begin{multline}
\nu(U(-p_1)U(-p_2)U(-p_3))\nu(V(q_1)V(q_2)U(p_3))\nu(V(-q_1)U(-p_2)V(-q_3)) \\ \nu(U(p_1)V(-q_2)V(-q_3))=-1,
\end{multline}
giving the required contradiction.

\section{An Operator Algebraic Perspective: D\"{o}ring's Theorem}

The most general form of the Kochen-Specker theorem is due to D\"{o}ring \cite{dor1} and
concerns valuations on a general von Neumann algebra $\mathcal{R}$ (viewed as a von Neumann subalgebra of $\mathcal{L(H)}$ for some separable Hilbert space $\mathcal{H}$). 

\begin{theorem} (D\"{o}ring)\label{th:dor}\\
Let $\mathcal{R}$ be a von Neumann algebra, with self-adjoint part $\mathcal{R}_{\rm sa}$, without type $I_1$ or $I_2$ summand. 
Then there is no $\nu : \mathcal{R}_{\rm sa} \to \mathbb{C}$ satisfying SPEC and FUNC.
\end{theorem}

D\"{o}ring actually provides a number of different Kochen-Specker-type theorems in \cite{dor1} which arise from two Gleason-type theorems.
The simplest, corresponding closely in its conclusion to the original Kochen-Specker result 
(in which $\mathcal{R} = M_3(\mathbb{R})$), and of which the situation arising from the Mermin system is an example, 
is a direct consequence of Gleason's theorem \cite{glea1}, which we recall
states that completely additive measures on the projection lattice $\mathcal{P(H)}$ are given by normal states and thus by density operators via the usual trace formula. The lack
of a dispersion-free (i.e., $\{0,1\}$-valued) state on the projection lattice is then enough to
rule out a valuation $\nu$. The form of Kochen-Specker theorem thus arising 
is valid on type $I$ factors. 

Remarkably, as D\"{o}ring demonstrates,
any non-abelian von Neumann algebra without summands of type $I_1$ and $I_2$ has enough structure to display
a Kochen-Specker contradiction. We refer to \cite{dor1} for a proof which relies on
a generalised version of Gleason's theorem (see \cite{chr1,yea1,yea2,mae1}; given also in \cite{ham1}) applicable to general von Neumann algebras (i.e., not necessarily factors, and certainly not of type $I$). Roughly speaking, this version
of Gleason's theorem shows that additive measures extend to states provided that the given
algebras have no type $I_2$ summand.

It is immediately obvious that the von Neumann algebra generated by
the state-independent Mermin proof falls within the remit of D\"{o}ring' theorem.
Indeed, it is of the simplest form: consider the ten observables $A_i$ appearing in Mermin's proof, and form $\A := \{A_i\}^{\prime \prime}$ (prime denoting commutant). Of course, $\A=\{ X_1, X_2, X_3, Z_1, Z_2, Z_3\}^{\prime \prime}$, and it is readily verified that $\A=M_2(\mathbb{C})\otimes M_2(\mathbb{C}) \otimes M_2(\mathbb{C}) \cong M_8(\mathbb{C})$. Clearly, $\A$ is a type $I_8$ factor and and hence there is no valuation on $\A_{\rm sa}$.

\section{Geometry: Root Systems, Weyl Chambers and the Emergence of $E_8$}

Mermin's proof connects to the Coxeter group  $E_8$ in two simple steps: one collects together the eigenstates defined by the contexts, and then a process of ``completion" is effected as described below. The final set turns out to be the root system of $E_8$, which can be verified
using Coxeter diagrams. Additionally, one can exploit the simple transitive action of $E_8$ on its set of chambers and get a description
of Mermin's proof in terms of galleries.  

We briefly review the mathematics involved and refer the reader to \cite{humphreys, nicolas} for details. A
{Coxeter matrix} indexed by $S$, with $S$ a finite set,  is a function 
$m_{i j}:S \times S \into \{1, 2, \cdots\} \cup \{+\infty\}$ satisfying
$m_{ii}=1$ and $m_{i j}=m_{ji} >1$ for $i\neq j$.  Associated to a Coxeter matrix is a
{Coxeter diagram}. Its set of nodes  is $S$ and the two nodes $i$ and $j$ are connected if $m_{i j}\geq 3$: we assign 0, 1, 2, or 3 edges between $i$ and $j$ when $m_{ij}$ is 2, 3, 4, or 6, respectively.   
The {Coxeter group} $W=W_S$ associated to a Coxeter matrix $m$ is the group with generators $s_i, \, i\in S$
and relations: (1) $s_i^2=1$ and (2)  $(s_is_j)^{m_{ij}}=1$ (the braid relations).  The pair $(W, S)$ is called a {Coxeter system}. The canonical geometric realisation of Coxeter groups uses root systems.
There, the elements of $S$ are represented by real reflections. 
A root system $\Phi \subset \R^n$  (with $n=card(S))$ is a finite set of vectors (roots) satisfying the two simple properties: 
\begin{enumerate}
	\item For all $\alpha \in \Phi$: $\lambda \alpha\in \Phi$ iff $\lambda=\pm 1$,
	\item For all $\alpha, \, \beta \in \Phi$: 
	 $$s_\alpha(\beta): = \beta - 2\frac{(\alpha, \beta)}{(\alpha, \alpha)}\, \alpha  \in \Phi.$$
\end{enumerate}
The pairing is the standard inner product on $\R^n$. The transformation $s_\alpha$ 
is the { reflection} with respect to the {hyperplane (wall)} $ \{x: (x, \alpha)=0\}$; it is an involution.  Reflections  $s_\alpha$  for $\alpha\in \Phi$ generate the group $W$ (more precisely, generate the Tits realisation of $W$ as a reflection group).   By property (2), it is a symmetry group of  $\Phi$ and $W$ is a subgroup of the group $\subset O(\R^n)$ of all
orthogonal transformations on $\R^n$. Hence,  $W(\Phi)$ preserves angles  and lengths.

\medskip

We return now to Mermin's proof and explain the emergence of the group  $E_8$. Each context in the proof yields a basis for the Euclidean space $\R^8$. We get  40 vectors in total, corresponding to the five bases. By reflections, this set of vectors is completed to
240 vectors $\alpha_i$ with $\alpha_{i+120} = - \alpha_i$ for all $1\leq i \leq 120$. Coxeter's diagram computation yields the following diagram,
\begin{figure}[!h]
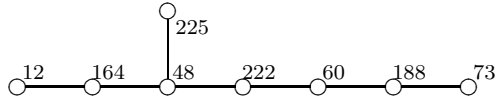

  \centering
$\xy
\POS (-10,0) ="z",
\POS (0,0) *\cir<3pt>{} ="a",
\POS (10,0) *\cir<3pt>{} ="b",
\POS (20,0) *\cir<3pt>{} ="c",
\POS (30,0) *\cir<3pt>{} ="d",
\POS (20,10) *\cir<3pt>{} ="e",
\POS (40, 0) *\cir<3pt>{} ="f",
\POS (50, 0) *\cir<3pt>{} ="g",
\POS (60, 0) *\cir<3pt>{} ="h",
\POS (70,0)  ="y",
\POS "a" \ar@{-}^<<{12} "b",
\POS "b" \ar@{-}^<<{164} "c",
\POS "c" \ar@{-}^<<{48} "d",
\POS "e" \ar@{-}^<<{225} "c",
\POS "d" \ar@{-}^<<{222} "f",
\POS "f" \ar@{-}^<<{60} "g",
\POS "g" \ar@{-}^<<{188} "h",
\POS "h" \ar@{}^<<{73} "y",
\POS "z" \ar@{}^<<<<{} "a",
\endxy$
\caption{Coxeter diagram of $E_8.$}
\end{figure}
 which uniquely identifies the Coxeter group $E_8$ (a fact about  finite irreducible  Coxeter groups). 

A Weyl chamber  of a root system $\Phi$ is the geometrical realisation of the generating set $S$;  it is the connected region of $\R^n$ bounded by the walls of the different  generating reflections in $S$.    
 A {gallery} is a sequence of chambers $S_i$ with each successive pair sharing
a face of codimension one. For instance, if $w=s_1\cdots s_n$, then the sequence $S, s_1S, s_1s_2S, \cdots,s_1s_2\cdots s_n S$ is a gallery. 
The Coxeter group $W$ acts transitively on its set of chambers; this will help us to translate  Mermin's proof from the level of roots to that of chambers. 
First, we fix a reference chamber $S_0$. 
Suppose $S_0$ is the chamber given by the diagram of Figure 1. 
The reflection $s_1$ (with respect to the first root)
 is then represented by the chamber
 
 \begin{center}
$
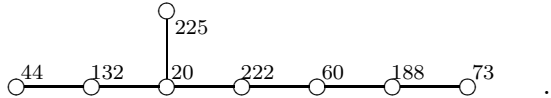
\xy
\POS (-10,0) ="z",
\POS (0,0) *\cir<3pt>{} ="a",
\POS (10,0) *\cir<3pt>{} ="b",
\POS (20,0) *\cir<3pt>{} ="c",
\POS (30,0) *\cir<3pt>{} ="d",
\POS (20,10) *\cir<3pt>{} ="e",
\POS (40, 0) *\cir<3pt>{} ="f",
\POS (50, 0) *\cir<3pt>{} ="g",
\POS (60, 0) *\cir<3pt>{} ="h",
\POS (70,0)  ="y",
\POS "a" \ar@{-}^<<{44} "b",
\POS "b" \ar@{-}^<<{132} "c",
\POS "c" \ar@{-}^<<{20} "d",
\POS "e" \ar@{-}^<<{225} "c",
\POS "d" \ar@{-}^<<{222} "f",
\POS "f" \ar@{-}^<<{60} "g",
\POS "g" \ar@{-}^<<{188} "h",
\POS "h" \ar@{}^<<{73} "y",
\POS "z" \ar@{}^<<<<{} "a",
\endxy.$
\end{center}
~\\
The second reflection $s_2$ is represented by

 \begin{center}
$
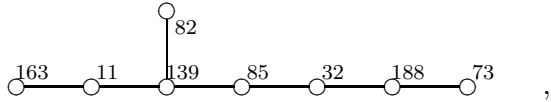
\xy
\POS (-10,0) ="z",
\POS (0,0) *\cir<3pt>{} ="a",
\POS (10,0) *\cir<3pt>{} ="b",
\POS (20,0) *\cir<3pt>{} ="c",
\POS (30,0) *\cir<3pt>{} ="d",
\POS (20,10) *\cir<3pt>{} ="e",
\POS (40, 0) *\cir<3pt>{} ="f",
\POS (50, 0) *\cir<3pt>{} ="g",
\POS (60, 0) *\cir<3pt>{} ="h",
\POS (70,0)  ="y",
\POS "a" \ar@{-}^<<{163} "b",
\POS "b" \ar@{-}^<<{11} "c",
\POS "c" \ar@{-}^<<{139} "d",
\POS "e" \ar@{-}^<<{82} "c",
\POS "d" \ar@{-}^<<{85} "f",
\POS "f" \ar@{-}^<<{32} "g",
\POS "g" \ar@{-}^<<{188} "h",
\POS "h" \ar@{}^<<{73} "y",
\POS "z" \ar@{}^<<<<{} "a",
\endxy,$
\end{center}
~\\
etc. The fact that $m_{1, 2}=2$ means that two galleries $S_0, S_1, s_2 S_1$ and $S_0, S_2, s_1 S_2$ intersect 
and define a cycle. 
Now, suppose the first root is picked (coloured black); then we have a valuation which assigns 1 to $s_1$ and 0 to $s_2$ (root 2 coloured white) and so on. This valuation also assigns 1 to $s_1s_2$. The gallery  $S_0, S_1, s_2 S_1$ is then coloured black. This is also valid for any gallery $S_0, S_1, s_i S_1$ with $(s_1s_i)^2=1$. The gallery $S_0, S_2$ is white, which prevents the whole cycle from being coloured black. 
For each context, we construct all such galleries. We then group them all together to get a centred network (all galleries start from $S_0$).
The translation of the colouring proof is now straightforward. 

\section{(Co-)presheaves and (Co-)sheaves}

The Kochen-Specker  theorem can be elegantly phrased in terms of functors from a category
to the category of sets. In this section we review three such formulations: the spectral presheaf
$\Sigma$ of Isham {\it et al.}, where the (state-independent) Kochen-Specker theorem concerns the\linebreak (non-)existence of points of $\Sigma$; the covariant approach of Heunen {\it et al.}, and the operational approach of Abramsky {\it et al.}

\subsection{Isham et al.}\label{subsec:isham}

The base category under consideration in the work of Isham, Butterfield, and Hamilton (e.g., \cite{ib3}) is 
$\mathcal{V(A)}$, with $\mathcal{A}$ a $C^*$-algebra, $\mathcal{V(A)}$ denoting the partially ordered set (poset) of unital abelian $C^*$-subalgebras, and elements of $\mathcal{V(A)}$
now referred to as contexts. The morphisms in $\mathcal{V(A)}$ are inclusion relations; $U \to V$
if and only if $U \subseteq V$ as abelian algebras (we will also use $\hookrightarrow$ for inclusion arrows).

A presheaf of particular relevance to the Kochen-Specker theorem is the spectral presheaf $\Sigma : \mathcal{V(A)} \to \mathsf{Set}$. For each $V \in \mathcal{V(A)}$,
\begin{itemize}
\item $\Sigma (V):= \{\lambda_V : \lambda_V \text{is a character of } V\}$ equipped with the topology of pointwise convergence, i.e., $\Sigma (V)$ is the Gelfand spectrum of $V$ (in the case of the Mermin proof, the topology is discrete);
\end{itemize}
on inclusions $U \hookrightarrow V$,
\begin{itemize}
\item $\Sigma (U \subseteq V) : \Sigma (V) \to \Sigma (U)$; $\lambda_V \mapsto \lambda_V |_U \equiv \lambda _U$, where the restriction is understood in the sense of function restriction.
\end{itemize}

The state-independent Kochen-Specker theorem concerns the existence of a global section (or \emph{point}) $\tau:1 \to \Sigma$ ($1$ is the terminal element). If such a section were to exist, for each $V$, $\tau_V : \{*\} \to \Sigma (V)$, and
for each $U \hookrightarrow V$, $\tau_V(*)(\lambda_V)=\lambda_V|_U \equiv \lambda_U$. In other words, we would have a valuation for each context giving rise to a global valuation.

The Mermin proof can be phrased in these terms when restricted to the poset $\mathcal{P}$. Let $\tau$ be such a putative global section.
Then, since $\tau_{V_i}(*)$ is a character of each abelian algebra $V_i$,

\begin{eqnarray}
\tau_{V_1}(*)(X_1X_2X_3)\tau_{V_1}(*)(X_1)\tau_{V_1}(*)(X_2)\tau_{V_1}(*)(X_3) = 1,\label{eq:1'}\\
\tau_{V_2}(*)(X_1Z_2Z_3)\tau_{V_2}(*)(X_1)\tau_{V_2}(*)(Z_2)\tau_{V_2}(*)(Z_3) = 1,\label{eq:2'}\\
\tau_{V_3}(*)(Z_1X_2Z_3)\tau_{V_3}(*)(Z_1) \tau_{V_3}(*) (X_2) \tau_{V_3}(*)(Z_3) = 1,\label{eq:3'}\\
\tau_{V_4}(*)(Z_1Z_2X_3)\tau_{V_4}(*)(Z_1)\tau_{V_4}(*)(Z_2) \tau_{V_4}(*)(X_3) = 1,\label{eq:4'}\\
\tau_{V_5}(*)(X_1X_2X_3)\tau_{V_5}(*)(X_1Z_2Z_3)  \tau_{V_5}(*) (Z_1X_2Z_3) \tau_{V_5}(*)(Z_1Z_2X_3)=-1\label{eq:5'}.
\end{eqnarray}

However, the choice of characters on each algebra is constrained by the requirement
that $\tau_{V_1}(*)(X_1)= \tau_{V_2}(*)(X_1)$, 
$\tau_{V_5}(*)(X_1X_2X_3) = \tau_{V_1}(*)(X_1X_2X_3)$, etc. After making these replacements in equations \eqref{eq:1'}--\eqref{eq:5'}, one again meets with a contradiction.

Given a unit vector $\varphi \in \mathbb{C}^8$ one can define a subpresheaf $\mathfrak{w}^{\varphi}$ of $\Sigma$ in which state-dependent proofs may be considered. Indeed, as demonstrated in
\cite{ldr1}, the GHZ state $\state{\Psi}$ gives rise to the  presheaf $\mathfrak{w}^{\Psi}$
which lacks a global section.

A primary motivation for Isham {\it et al.} for constructing $\Sigma$ and $\mathfrak{w}^{\varphi}$ as they do is that they may be viewed as objects in the topos 
$\mathsf{Set}^{\mathcal{V(A)}}$, i.e., the topos of all set-valued presheaves on $\mathcal{V(A)}$. A topos
has a multivalued ``truth object" provided by the subobject classifier; Isham and D\"{o}ring \cite{ido1}
then argue for a connection between the quantum probabilities given by the Born rule and 
the truth values in $\mathsf{Set}^{\mathcal{V(A)}}$. Furthermore, they argue, the relaxation 
of a Boolean truth system to a more general, many-valued intuitionistic one paves the way for 
a ``neo-realist" interpretation of quantum theory. 

\subsection{Heunen-Landsman-Spitters}

Heunen, Landsman, and Spitters \cite{hls1} have provided another topos-theoretic view on the 
Kochen-Specker theorem. With $\mathcal{V(A)}$ denoting the poset of unital, abelian $C^*$-subalgebras of 
some fixed $C^*$-algebra $\mathcal{A}$, they construct the topos $\mathcal{T(A)}$---the topos of all \emph{covariant} set-valued functors on $\mathcal{V(A)}$.

The point is to \emph{internalise} the (in general) non-abelian algebra $\mathcal{A}$, yielding
$\underline{\mathcal{A}}$ internal to $\mathcal{T(A)}$ (in this section we follow the convention of the authors of \cite{hls1} in underlining internal entities) which \emph{is} abelian, and is referred to
as the \emph{Bohrification} of $\mathcal{A}$.

$\underline{\mathcal{A}}$ is the ``tautological functor", defined on objects by
\begin{equation}
\underline{\mathcal{A}}(C)=C,
\end{equation}
and on any arrow $C \subseteq D$, $\underline{\mathcal{A}}(C) \hookrightarrow \underline{\mathcal{A}}(D)$. The $C^*$-algebra $\mathcal{A}$ is a (not necessarily commutative) $C^*$-algebra in the topos $\mathsf{Set}$, whereas, as proved in \cite{hls1}, $\underline{\mathcal{A}}$ is a commutative $C^*$-algebra in the ``universe of discourse" provided by 
$\mathcal{T(A)}$.

The covariant ``version" of the spectral presheaf $\Sigma$ from Section \ref{subsec:isham}, which will be denoted $\underline{\Sigma}$, gives the internal Gelfand spectrum 
$\underline{\Sigma}(\underline{\mathcal{A}})$ of the internal abelian algebra $\underline{\mathcal{A}}$. $\underline{\Sigma}(\underline{\mathcal{A}})$ is an internal locale;
the points of a general locale $X$ in a topos are the frame maps $\mathcal{O}(X) \to \Omega$, with $\Omega$ being the subobject classifier of the topos in question. As shown in \cite{hls1}, the Kochen-Specker theorem now reads:

\begin{theorem}\label{th:hls1} (Heunen-Landsman-Spitters)\\
Let $\mathcal{H}$ be a Hilbert space for which $dim \mathcal{H} \geq 3$ and let $\mathcal{A} = \mathcal{L(H)}$. Then $\underline{\Sigma}(\underline{\mathcal{A}})$ has no points.
\end{theorem}

The proof is roughly as follows (see \cite{hls1}, Theorem 6, and the accompanying proof):
internally, a point $\underline{\rho}: \underline{*}\rightarrow \underline{\Sigma}$ of the locale 
$\underline{\Sigma}$ can be combined with a self-adjoint operator $a\in \underline{A}_{\mathrm{sa}}$ through its Gelfand transform  
$\hat a:\underline{\Sigma}\rightarrow \underline{\mathbb R}$ to give a point 
$\hat a\circ \underline{\rho}: \underline{*}\rightarrow \underline{\mathbb R}$, the latter denoting the locale of Dedekind reals, again viewed internally. This defines an internal multiplicative linear map (natural transformation) $v_\rho:   \underline{A}_{\mathrm{sa}} \rightarrow {\mathrm{Pt}( \underline{\mathbb R})}$ with components $v_\rho(V):  \underline{A}_{\mathrm{sa}}(V) \rightarrow {\mathrm{Pt}( \underline{\mathbb R})}(V)$, which coincides externally with $C_{\mathrm{sa}} \to \mathbb{R}$, and
 is precisely the valuation that is ruled
out by the Kochen-Specker theorem \cite{KS1}.

With regard to the Mermin system, replacing $\mathcal{V(A)}$ by $\mathcal P$ generated by the five contexts involved, we may view the (non-commutative) $C^*$-algebra $\mathcal{A_P}$ again
as an internal commutative $C^*$-algebra $\underline{\mathcal{A_P}}$ in the topos
$Sets^P \equiv \mathcal{T}(\mathcal{A_P})$ with associated internal locale 
$\underline{\Sigma}(\underline{\mathcal{A_P}})$. That this locale has no points follows from
considering valuations $v_\rho(V)$ for the abelian algebras of each context, yielding
again the insoluble set of equations \eqref{eq:1'}--\eqref{eq:5'}, with $v_\rho(V)$ coinciding with $\tau(*)_V$.

\subsection{Abramsky et al.}\label{subsec:Abram}

Abramsky and Brandenburger \cite{abram1} have presented an operational, sheaf-theoretic description of contextuality, and Abramsky, with Mansfield and Barbosa, have developed cohomological techniques \cite{abrampara, abram2} for identifying its presence. 
Their approach is independent of the Hilbert space formalism, thereby doing away with 
SPEC and FUNC and instead focussing directly on the (im-)possibility of a global section
for a compatible family of no-signalling distributions.

Following Abramsky and Brandenburger, we refer to a finite collection of observables $X$, understood as a topological space with the discrete topology
and distinguished open sets (the \emph{contexts} here) $\{C_i \} \equiv \mathcal{M}; \bigcup C_i = X$ as a \emph{measurement cover}. Each $C_i$ is a maximal set of observables which may be measured jointly.
The outcomes of a measurement of any observable $A \in X$ are given by the set $O$ and, therefore, under the assumption that all observables have the same outcome sets,
for a measurement of $n$ observables in $X$, outcomes are in $O \times O \times ... \times O \equiv O^n$. The \emph{event sheaf} $\mathcal{E}$ 
is defined
by $\mathcal{E}(U):= O^U \equiv \{ f:U \to O\}$ for $U \subset X$, i.e., the collection of assignments for observables in $U$. Elements of $\mathcal{E}^U$,
or \emph{sections} above $U$, are therefore viewed as possible assignments of observables in
$U$. An \emph{empirical model} $e$ specifies joint probability distributions $e(C)$ over assignments
in $\mathcal{E}(C)$, where $C \in \mathcal{M}$, to be compatible in the sense of the no-signalling principle (or, equivalently, the sheaf condition: see \cite{abram1}).
The \emph{support} presheaf $S_e$
 of $e$ specifies a subpresheaf of $\mathcal{E}$  defined by $S_e(U)=\{s \in \mathcal{E}(U)~:~s \in \text{supp}e_U\}$.

Abramsky and Brandeburger \cite{abram1} identify three levels of contextuality/non-locality arranged in a proper hierarchy---probabilistic non-locality, possibilistic non-locality, and strong contextuality---in which the state-dependent Mermin system, and indeed all AvN arguments, occupies the strongest level of strong contextuality. The next level
down corresponds to the situation where outcome probabilities can be neglected and all the matters is what is \emph{possible}, i.e., on the level of supports; the Hardy model occupies this level and is \emph{not} strongly contextual. In turn, the Bell model is probabilistically non-local but \emph{not} possibilistically non-local, thus showing that each level in the hierarchy can be realised by quantum mechanics. It should be noted, however, that quantum mechanics is not a necessary requirement---for example, Podolosky-Rosen boxes are also strongly contextual \cite{abram1}.

\begin{definition}
The model $e$ is called \emph{strongly contextual} if $S_e(X) = \emptyset$.
\end{definition}
As shown in \cite{abram1}, the \emph{GHZ models} (for at least three parties)   are strongly contextual; We also note that all AvN arguments are strongly contextual. To exhibit strong contextuality in our familiar setting 
we turn our attention to the Mermin state-dependent proof, with the following table:

\begin{center}
    \begin{tabular}{| l | l | l | l | l | l | l | l | l |}
    \hline
    & $+++$ & $++-$ & $+-+$ & $+--$ & $-++$ & $-+-$ & $--+$ & $---$  \\ \hline
    $C_1 = X_1, X_2, X_3$ & $1$ & $0$ & $0$ & $1$ & $0$ & $1$ & $1$ & $0$ \\ \hline
    $C_2=X_1, Y_2, Y_3$ & $0$ & $1$ & $1$ & $0$ & $1$ & $0$ & $0$ & $1$ \\ \hline
    $C_3=Y_1, X_2, Y_3$ & $0$ & $1$ & $1$ & $0$ & $1$ & $0$ & $0$ & $1$ \\
    \hline
    $C_4 = Y_1, Y_2, X_3$ & $0$ & $1$ & $1$ & $0$ & $1$ & $0$ & $0$ & $1$ \\
    \hline
    \end{tabular}
\end{center}

There are four ``potential" global sections compatible with the assignments for $C_2$, $C_3$, and $C_4$,
namely
\begin{eqnarray}
s_1 : (X_1, X_2, X_3, Y_1, Y_2, Y_3) \mapsto (++-++-),\\
s_2 : (X_1, X_2, X_3, Y_1, Y_2, Y_3) \mapsto (+-++-+),\\
s_3 : (X_1, X_2, X_3, Y_1, Y_2, Y_3) \mapsto (-++-++),\\
s_4 : (X_1, X_2, X_3, Y_1, Y_2, Y_3) \mapsto (------).
\end{eqnarray}

As can be seen from the table, none of these are compatible with the constraints imposed
by $C_1$, thus yielding the required contradiction.

To conclude this section, we briefly remark that Abramsky {\it et al.} \cite{abram2, abrampara}
have observed cohomological witnesses for contextuality, including strong contextuality, 
by imposing a ring structure on the outcome spaces. These tools are strong enough
to identify the presence of an AvN argument, for instance. In general, their cohomological
construction yields sufficient conditions under which a Kochen-Specker contradiction occurs;
unfortunately, these are sometimes not necessary \cite{man1}.


%

\section{Algebraic Geometry}
It was observed in subsection \ref{subsec:isham} that the topology on $\Sigma(V)$ is discrete
in the Mermin example. If one views the discrete topology as the Zariski topology, then 
one can phrase the contradiction in the framework of algebraic geometry. For that, let $A$  be a commutative ring with identity. It will be, for us, of the form
$$\Z_2[i, j, \cdots, k]/ \p := \{\mbox {polynomials in }  i, j, \cdots, k  \mod \p\},$$
i.e.,  a coordinate ring of an affine variety $V(\p)\subset \Z_2^n$. The ideal  $\p$, 
in our context, represents the algebraic constraints involved in Mermin's type of proofs.  The discreteness of 
the situation allows the replacement of the Gelfand spectrum by the prime spectrum of an algebraic variety to be defined next.  
Specifically, the spectrum $\Sigma(\mathcal A)$ of the commutative algebra $\mathcal A$ is replaced by
the prime spectrum of a ring $A$ associated to $\mathcal A$. 

\begin{definition}[Prime spectrum of a ring]
	The prime spectrum of a ring $A$ is the set $\mathrm{PSpec}\, A$ of all prime ideals in $A$. 
\end{definition}
In our case of coordinate rings, $\mathrm{PSpec}\,  A$ has two types of elements: maximal ideals and non-maximal (but prime) ideals. 
The first type corresponds to points in the affine space $\Z^n$, whereas prime ideals which are not maximal are irreducible subvarieties
of our variety.  Each element $f\in A$ defines a function on the prime spectrum of $A$.  Let $\p$ be  an element of $\mathrm{PSpec}\,  A$. 
We denote by $\kappa(\p)$ the quotient field   $A/\p$, called the {residue} field 
of  $\mathrm{PSpec}\,  A$ at $\p$. We can define the value of $f$ at $\p$ to be the image of $f$ via the ring homomorphism
$
	v_\p:A\into  \kappa(\p).
$
If $A$ is a coordinate ring of an affine variety over an algebraically closed field $K$ and $\p$ is the maximal 
ideal corresponding to a point of the variety, then $\kappa(\p) =K$ and $f(\p)$ is just the valuation of $f$
at this point.

\medskip

 The goal of the remainder of this section is to  explain contextuality of Mermin's system in the language of algebraic geometry we have just reviewed. 
 Notation is as defined in the previous section. We ``identify" each  set $\mathcal S_e(C)$, where $C\in \mathcal P$ and $\mathcal P$ is the poset generated by $\mathcal M$, 
as an affine variety, and local sections in $\mathcal S_e(C)$  as maximal (ring) ideals  
in $\mathrm{PSpec}\,  A_C$. 
This yields for each $C$ a coordinate ring $A_C$. The set of rings $A_C$, which are partially ordered by inclusion,  define the poset ${\bf CoordRings}(\mathcal P)$. 
We can define the functor 
  \begin{eqnarray}\nonumber
 	 	   \mathcal F  : & {\bf CoordRings}(\mathcal P) & \longrightarrow \Sets\\
			& A & \longmapsto  \left\{ v_\p,  \quad \p\in \mathrm{PSpec}\,  A\right\}.
 \end{eqnarray} 
For Mermin-type Kochen-Specker proofs, the functor $\mathcal F$ has no point. We explain this in the Mermin example, which should be enough to convince
the reader of the validity of the latter statement.  
For the first context, we have the coordinate ring 
$
	A_1 := \Z_2 \left[ s_1, s_2, s_3\right]/\left(s_1s_2s_3-1\right)
$ 
where, for instance, the local section $(s_1, s_2, s_3) \mapsto (-1, -1, 1)$ {of $\mathcal S_e$ at the first context} defines a maximal
ideal $\p =\left(s_1+1, s_2+1, s_3-1 \right) \in \mathrm{PSpec}\, A_1$.  For the second context,  we have the coordinate ring
$
	A_2 := \Z_2 \left[ s_1, \tilde s_2, \tilde s_3\right]/\left( s_1\tilde s_2\tilde s_3+1\right),
$
and similarly for the two other maximal contexts. 
For the context $\{x_1\}$, the coordinate ring
is
$
	A_{1, 2} := \Z_2 \left[ s_1\right] 
$,
with two proper maximal ideals $(s_1-1)$ and $(s_1+1)$; i.e., the corresponding affine variety is the whole line $\Z_2$.

 For each ring $A_i$ and each maximal ideal $\p \in  \mathrm{PSpec}\, A_i$,   the local valuation
 $v_\p:A_i \into \kappa(\p)= A_i/\p$ takes $x\in A_i$ and returns  $ v_\p(x)  = x \mod \p$. 
In the case of  $A_1$ and $\p = \left(s_1+1, s_2+1, s_3-1 \right)$,  we have
 \begin{eqnarray}
 	v_\p\left( x_1(s_1, s_2, s_3)\right) &=& v_\p(s_1) = s_1 \mod \p =-1, \\\nonumber
	v_\p\left( x_2(s_1, s_2, s_3)\right) &=& v_\p(s_2) = s_2 \mod \p =-1, \\\nonumber
	v_\p\left( x_3(s_1, s_2, s_3)\right) &=&  v_\p(s_3) = s_3 \mod \p =~~1.  \\\nonumber
 \end{eqnarray}
 The functor $\mathcal F$ has no points for GHZ3:  for any maximal ideal $\p_1\in A_1$, one has
  \begin{equation}
  	v_{\p_1} \left(s_1 s_2 s_3 \right) =  v_{\p_1}(s_1) \,  v_{\p_1}(s_2) \,v_{\p_1}(s_3) =1,
  \end{equation}
  which is nothing but the equation (\ref{eq:mermcon}). For the three other maximal rings, we have similar equations
valid for all maximal ideals $\p_i\in \Spec A_i$. The system is inconsistent and the functor $ \mathcal F$ is pointless. 
  
\section{Discussion and Outlook}

As has been presented, there is a remarkable variety 
and mathematical richness surrounding
Kochen-Specker proofs, much of which is exhibited by Mermin's simple system. We saw that the collection of observables given in that example, along with a given state where appropriate,
yield naturally combinatoric contradictions in the form of insoluble equations, impossible colourings, and so on. Mermin's system, as Clifton showed, could also be adapted to the case of position and momentum observables. 

D\"{o}ring provided a non-combinatorial proof of the Kochen-Specker theorem in the language 
of von Neumann algebras, thereby providing a genuine generalisation. We then provided
some geometric insight on the Mermin proof through Coxeter groups and their root systems,
and observed that $E_8$ arises naturally in this context. The presheaf perspectives presented
then provide another way of phrasing Kochen-Specker-type contradictions---three such approaches were given. Finally, an
algebraic geometric framework was outlined. It was observed that in the finite scenario,
the Gelfand spectrum of an abelian algebra can be replaced by the prime spectrum of an algebraic 
variety, and the statement that the spectral presheaf of Isham and Butterfield has no
global section ``translates" into the statement that the prime spectrum functor has no global section. This algebraic geometric approach seems to warrant further exploration.

We only briefly touched upon the role played by cohomology in the Kochen-Specker theorem. 
We believe this to be an area for much further work, particularly when phrased in the language 
of toposes \cite{mord1}.  Let  
$\mathcal E$ be  the topos ${\sf Sets}^{\mathcal P^{op}}$ and
$\mathcal{AB}\left (\mathcal E\right)$ the category of of abelian group objects, i.e., presheaves from the poset $\mathcal P$ into the category $\mathcal{AB}$ of abelian groups. The Giraud axiom for generators implies that the abelian category $\mathcal{AB}(\mathcal E)$ has enough injectives. The global sections functor $\Gamma: \mathcal E\rightarrow {\sf Sets}$ induces a functor (again denoted) $\Gamma: \mathcal{AB}(\mathcal E)\rightarrow \mathcal{AB},$ which is left exact and preserves injectives. For any abelian group
object $A\in \mathcal E,$ the cohomology groups $H^n(\mathcal E, A)$ are defined as the right derived functors of $\Gamma$.
Now, the functors
$\Sigma$ and $S_e$ are in $\mathcal{AB}\left (\mathcal E\right)$ in the case of GHZ3; i.e., they are functors of vector spaces. Mermin's proof translates into the functors $H^n$, for both $\Sigma$ and $S_e$,  being all equal to the zero functor. 

%
%


\section*{Acknowledgements}
Thanks are due to Shane Mansfield for useful discussions and a careful reading of this manuscript.
We also thank Marko Bucyk for his help. Leon Loveridge acknowledges funding from the grant \emph{Quantum Mathematics and Computation}.


\end{document}